\documentclass[
    ,final            
  ]
  {aipproc}
\usepackage{amssymb,graphicx,amsmath,axodraw}
\layoutstyle{6x9}

\begin{document}
\begin{flushright}
PREPRINT ULB-TH/06-19\end{flushright}

\title{About Mass, CP  and Extra Dimensions}


\author{J.-M. Fr\`{e}re}{
  address={Physique Th\'{e}orique CP 225, ULB, B-1050 Bruxelles Belgium}
}

\begin{abstract}
 We discuss the notion of mass, mostly for fermions, and its
 relation to the breaking of CP invariance, the natural symmetry of
 gauge interactions.  In a first model,
 we show how compactification on a Vortex in 2 extra dimensions
 leads to a replication of generations in 3+1, with challenging mass
 patterns, and testable consequences in flavour-changing neutral
 currents (family-number conserving), both at low energies and at
 future colliders.
 In different model, we show how CP violation can result from
 compactification from 4+1 to 3+1 dimensions.
\end{abstract}

\maketitle


\section{Introduction}
  While gauge boson masses are intimately linked to broken local symmetry,
  no such logical connection occurs in general for fermions. In the Standard
  Model, at least some of the mass must be linked to symmetry breaking, but
  this is due to the fact that left- and right-handed fermions are not in
  the same representations. Hypothetical vectorlike fermions could indeed have masses
  without symmetry breaking. Another situation is known for the effective,
  sometimes called "constituant" mass. It is usually interpreted as a result of
  strong interactions leading to a breaking of chiral symmetry, and used as
  a model for "dynamical symmetry breaking".
  Quite another possible origin for masses is from extra dimensions. Here
  the problem is almost the opposite. If the scales associated with extra
  dimensions are heavy, it is now a matter of preserving some light states.
  This is often realized through some form of localisation on topological
  singularities: domain walls in one, or vortices for two extradimensions.

  In this presentation, we review two topics relating fermion masses
  and extra dimensions. The first aspect may seem at first a variant
  of fermion "multilocalisation", \cite{overlaps} leading
  to the generation of mass
  patterns from overlap of wave functions in extra dimensions. The
  scheme considered here is however much more constrained and
  deterministic, as the fermions are all localized in the same way
  on a unique topological singularity, and the overlap functions are
  precisely determined by the dynamics of this singularity. This
  work is based on the series of papers
  \cite{Frere:2000dc},\cite{Frere:2001ug},\cite{Frere:2003yv},
  \cite{Frere:2003ye},with predictions for colliders in
  \cite{Frere:2004yu}and recent conjectures about a light Brout-Englert-Higgs
  particle in \cite{Libanov:2005mv}.

  The second part deals with another fundamental
  question, namely: "How does CP (or T) violation enter in a
  fundamental theory?". The main point is that CP is the natural
  symmetry of gauge interactions, and therefore attempts at unifying
  scalar couplings (effective or not) with gauge interactions pose
  the problem of breaking CP symmetry. We prove that this can occur
  through dimensional reduction, by the inclusion of "Hosotani
  loops" (in this case, the equivalent of Wilson loops, but looping
  around the extra dimension). In a simple example, we generate CP
  violation from a 4+1 theory with only real couplings. This work is
  based on reference \cite{Cosme:2002zv}, and its generalization to
  include grand unified structures \cite{Cosme:2003cq}.

\section{Three Families in One}
In this section, we describe in general terms a model evolved in
collaboration with Serguey Troitsky and Maxim Libanov. Full
mathematical details can be found in the original papers, and we
will here mainly list the salient results.

It may still be useful to start with a little history of the model.
It was initially introduced by S. Troitsky and M. Libanov in a
different form, possibly lighter in field content, but with only
approximate symmetry. \cite{Libanov:2000uf}, using a vortex of
winding number 3. The following version, still formulated for flat
extra dimensions, simulated this vortex with a scalar field of
winding number one (with cubic coupling to fermions, and avoided any
explicit breaking of symmetry by introducing supplementary scalar
fields.\cite{Frere:2000dc}

In both cases, the first rationale for the use of a topological
defect (the vortex) was to confine the fermions to a small region of
the extra-dimensional space, taken otherwise to be flat and
infinite. The vortex being built out of a scalar field $\Phi$ and an
auxiliary gauge field $A^{\mu}$ (both unrelated to the Standard
Model fields, these fields don't enter directly the currently
observable phenomenology), it is possible to apply the Index
theorem, and to conclude that for an effective winding number $n$
exactly $n$ massless chiral fermion modes persist in the remaining
3+1 dimensions for each fermion field coupling to the 6-d structure.

We have thus developed the phenomenology from this onset, but soon
realized that the flat and unbounded extra dimensions represented a
liability, with the need of confining the observed gauge fields,
which otherwise could couple to fermionic modes outside the
structure. A more involved version, largely with the same
characteristics, was thus developed with the extra dimensions now
assuming spherical topology.

Why maintain the topological singularity?

Part of the answer lies of course in the counting of light (massless
at compactification scale) fermion modes, and the resulting
replication as an origin to the very notion of particle families.

Obviously, the number of light fermions is not a direct prediction
of the model in its present state, as we have to put in by hand the
winding number, but the mass hierarchy of fermions, and a number of
phenomenological implications are quite generic. We will detail them
somewhat now.

\subsection{The Model - fermion wave functions and mass hierarchies}
While the latest formulation has the extra dimensions on a sphere,
we will for pedagogical reasons describe the situation for a plane.
The outcome in terms of fermion wave functions, selection rules and
masses is very similar. For the plane, the extra variables are
chosen as $(r,\varphi)$, while for the sphere (of radius $R$) we use
$(R \theta, \varphi)$. Assuming the above-mentioned vortex structure
is described by a field $\Phi$ such that
\begin{equation}\label{background}
    \Phi(x,r,\varphi) = \Phi(x) \ {\Phi}_1(r) \ e^{i \varphi}
\end{equation}
This of course has only winding number 1, but the winding number 3
structure is achieved by coupling all  the fermion fields according
to:
\begin{equation}\label{winding3}
    \overline{\Psi} \Phi^3 \Psi
\end{equation}

This coupling may look surprising, as it is obviously
non-renormalisable in 4 dimensions. We are not considering
renormalisability of the theory here, as the 6 dimensional context
is most likely an effective model, but even in this case it might be
suitable that the 4-dim reduction be renormalisable. In this
context, the reason we proceed as above is mostly simplicity: we
could obviously replace this structure with a winding-3 solution for
a fundamental field $\widetilde{\Phi}$, and find similar solutions
for the couplings which will follow. This would however
unnecessarily clutter the discussion, and we opt at this point for
the simplicity of the field content.

The Yukawa coupling in equation \eqref{winding3} is only responsible
for linking the fermions to the vortex. For each type of fermion
field $\Psi_T$ introduced. Here the type $T$ represents  the various
quarks, $U_R,D_R, Q_L=(U_L,D_L)$ or the leptons  and the indices
$L,R$ refer to the 4-dim chirality of the associated particles after
dimensional reduction. Only one field is introduced to represent all
the "up" quarks, for instance.  The vortex structure leads, through
the index theorem to 3 massless localized chiral solutions in 3+1
dimensions, which we will associate with the family replication.
Typically, we have

\begin{equation}\label{fermionfamilies}
     {\Psi_T}_n\sim {f_T}_n(r) e^{i (3-n) \varphi}
\end{equation}

What is important here is that each "family" of solutions has a
different "winding" behaviour in the variable $\varphi$. The
labeling of the modes is quite arbitrary; it was chosen here so that
in the simplest case the larger mass is associated to the third
generation. Rather than analytic expressions, we give a pictorial
representation of the radial dependence of the corresponding
functions ${f_T}_n(r)$.
\begin{figure}
  \includegraphics[height=.2\textheight]{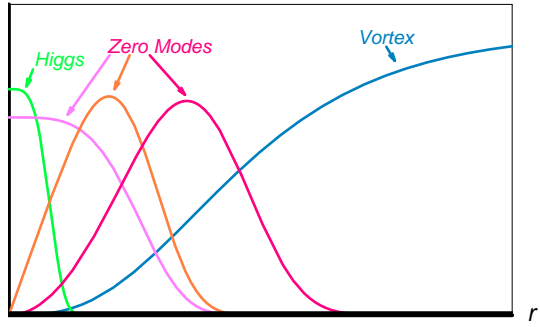}
  \caption{Profiles of the relevant wave functions in the radial coordinate}
  \label{all}
\end{figure}
The fermionic mode 3 is non-vanishing at $r=0$ , with the modes
behaving near the origin as:
\begin{equation}\label{power}
{f_T}_n(r) \sim r^{3-n} \ \; \ \  r\rightarrow 0
\end{equation}

We have thus realized a situation similar to, but much more
constrained than the multilocalisation in 5 dimensions. Here, we
don't have the freedom to place our various families at arbitrary
spacing, but are faced instead with non-trivial overlaps fixed by
the vortex properties.

This far, we only dealt with the "background" of the problem, and
generated essentially massless states in 3+1 dimensions. Now comes
the time to generate the "usual" fermion masses, and to break the
$SU(3) X SU(2) X U(1)$ symmetry. This is done in the usual way,
through the Brout-Englert-Higgs mechanism, at the price of
introducing a scalar doublet, which we will name $H$. In fact, for
the purpose of separating the various quantum numbers, we write,
instead of the usual coupling:

\begin{equation}\label{yukawa}
   { \cal L}_{Yukawa }= \lambda_U \Psi_{U_R}  H^\dagger X \Psi_{Q_L}
    + \epsilon \lambda_U \Psi_{U_R} \ H^\dagger \Phi \Psi_{Q_L} +
    h.c.
\end{equation}

where $\Phi$ is the background field already described, while $X$ is
a singlet with non-vanishing value at $r=0$ and winding number 0.
The reason for introducing these couplings and $X$ itself, is to
simplify the breaking structure; it would be perfectly possible to
introduce single fields to replace the combinations $H \Phi$ and $H
X$. Obviously, similar terms obtain for the other fermion types.

The $H$ field also couples to $\Phi$, and this results in a
non-trivial profile, centered at the origin, and exemplified in
figure\ref{all}. Note that  $H^\dagger X$ and $H^\dagger \varphi$
\begin{eqnarray} \label{higgswinding}
  H^\dagger X & \sim & e^{i\ 0\ \varphi} \\
  H^\dagger \varphi& \sim & e^{i \varphi}
\end{eqnarray}
\subsection{Mixing of families and alternatives}
The reduction to 4 dimensions involves integration over $\varphi$
and $r$ (or $\theta = r/R$ in the case of a sphere) and generates
the mass terms (we just give the "up" quark as an example,and
neglect the term in $\epsilon$ for the time being):
\begin{equation}\label{masses}
    m_{U,ij}= \lambda \int dr \int d\varphi f_{{U_R},i}^\dag f_{{Q_L},j} \ X(r) \ H(r)
    e^{i(j-i) \varphi}
\end{equation}
Clearly, the integration over $\varphi$ guarantees that only
diagonal entries occur, and the ratio of the masses is given by the
overlap between the fermion and scalar wave functions in the
coordinate $r$. If, for instance, the profile of the scalar
combination is concentrated close to the origin (as suggested in
figure \ref{all}), we get $ m_3 >> m_2 >> m_1 $ (more on this
later).

Mixing is however needed, and it can be introduced by the term
proportional to $ H^\dagger \Phi $ in equation \eqref{yukawa}. This
results in a mass matrix

\begin{equation}\label{massmatrix}
    m_{U,ij}=\left(
               \begin{array}{ccc}
                 m_1 & \epsilon \ m_{12} & 0 \\
                 0 & m_2 & \epsilon \ m_{23} \\
                 0 & 0 & m_3 \\
               \end{array}
             \right)
\end{equation}

The mass eigenstates are obtained by diagonalization, but the
Kobayashi-Maskawa matrix itself finds its origin in the differential
rotation of the "up" and "down" quarks. That these rotations are
different can result from different choices of $\lambda , \epsilon$.

The possibility also exists to use completely different patterns,
for instance choosing the $ H^\dagger \Phi$ contribution as leading
(which we are free to do, since the Yukawa couplings are arbitrary)
leads to a completely off-diagonal mass matrix, a feature which can
come in handy when dealing with the large rotations between
neutrinos and charged leptons.

\subsection{Family number near-conservation and Kaluza-Klein modes; Experimental constraints}
We turn now to the gauge interactions, and to their effect in 3+1
diemnsions. As usual, we will need to face a number of Kaluza-Klein
excitations for each gauge boson introduced. In the present case,
each such "tower" carries a double label, since it must refer both
to radial and angular excitations. Generically, ($A$ stands for any
gauge boson, $\gamma$, W, Z, gluons):
\begin{equation}\label{vectors}
    A_{lm}(r,\varphi )=a_{lm}(r){\rm e}^{im\varphi }
\end{equation}

The lowest state (massless before symmetry breaking) has a flat
profile in both variables, which guarantees charge universality on
one hand, and diagonal couplings (even in the mass basis) for the
neutral bosons. Mass relations between the $lm$ modes depend on the
topology of the model.

We will focus in this section on a particular feature of our
approach, namely the presence of flavour-changing neutral currents,
but with approximate conservation of "family number". Indeed, even
if we neglect the Kobayashi-Maskawa mixing, the $m \neq 0$ modes
generate transitions between families. Winding number ($\varphi$
dependence) acts as an effective "family number", which is conserved
at both vertices where an excitation $A_{lm}$ connects to the
fermions. For instance, the $Z$ boson excitations $ Z' \sim e^{\pm
i\varphi}$, allow the following transitions:
\begin{eqnarray}
  s+\overline{d}\Rightarrow  Z'\Rightarrow  s+\overline{d},\\
s+\overline{d}\Rightarrow  Z'\Rightarrow \mu+\overline{e},\\
s+\overline{d}\Rightarrow  Z'\Rightarrow\tau+\overline{\mu}
\end{eqnarray}

As discussed in \cite{Frere:2003ye}, this leads to powefull
constraints. In particular, the second process above induces the
transition $K\Rightarrow \mu+\overline{e}$, which is severely
constrained experimentally. The strength of the interaction is given
by the usual gauge coupling constant, the mass of the $Z'$
excitation, and an extra factor due to the overlap of the wave
functions for fermions of the first and second generation, and the
radial dependence of the $Z'$: $\kappa_{1,2} = \int dr \
f_{{U_R},1}^\dag \ f_{{Q_L},2} \ z_{0,1}(r)$, leading to the limit:
\begin{equation}\label{limit}
    M_{Z'}\geq \kappa _{12} \cdot 100 {\rm TeV}
\end{equation}

Future tests of this model will benefit much from currently planned
precision lepton flavour violation experiments at moderate energies,
as well as collider experiments trying to produce directly the
Kaluza-Klein excitations.

\subsection{Expectations at colliders}
We expect mainly two types manifestations at colliders:
\begin{itemize}
    \item the Brout-Englert-Higgs boson mass
    \item production of the Kaluza-Klein excitations.
\end{itemize}

The first issue is tackled in \cite{Libanov:2005mv}, where a
connection between the mass of $H$ and the extension of the scalar
wave function in the extra coordinate $r$. As already specified,
large mass ratios $m_3/m_2, m_2/m_1$ require that the wave function
be quite concentrated close to the origin.  For the simplest scheme
discussed above, \cite{Libanov:2005mv} gets an upper bound close to
the current limit.

The second issue deals with direct production of Kaluza-Klein
excitations. As seen from equation \eqref{limit}, there is here a
trade-off between the overlap factor and the actual mass of the new
gauge boson. Constraints from K meson decays put the mass range
above in the (unlikely) case that the overlap factor $\kappa$ is
close to 1. This is clearly out of reach of LHC, but a smaller
overlap may bring the scale down considerably, albeit at the cost of
production efficiency (thus requiring the full luminosity for
detection). We have studied this production as a function of
$\kappa$, assuming the bound on $M$ to be saturated in each case; in
\cite{Frere:2004yu}, we see that for an integrated luminosity of
$100 fb^{-1}$ and $\sqrt{s}=14 TeV$, a few events can be expected if
$M<3 TeV$ in the $\mu^+ e^-$ final mode (the charge-conjugate
channel is less sensitive). Note however that this is an
exceptionally clean final state to look for!

Considerably more events are expected if we deal with the
Kaluza-Klein excitations of the gluons instead, but there, even if
the family-number near conservation still obtains, the extraction of
the signal will be considerably more difficult, and detailed
modeling is requested.

\section{CP violation from extra dimensions}
\subsection{A toy model}
The motivation for this work was given already in the introduction:
if a fundamental theory has all couplings related to gauge
interactions, how can it violate the natural symmetry of those,
i.e., CP ? One possibility is to have \emph{spontaneous} CP
violation, involving several, relatively complex vacuum expectation
values. This is certainly possible, requires at least 3 such vev's,
but can appear somewhat contrived. We look here for a more
structural approach.

We start with a simple example, showing how a complex mass matrix
can arise from a purely real Lagrangian in 5+1 dimensions.

\subsubsection{P,CP and CPT in extra dimensions} Basically two
choices are possible for the definition of Parity: either flip all
the spatial coordinates, (central inversion), or only one (specular
reflection). For an odd number of spatial dimensions, they are
equivalent (up to a rotation), but not for an even number. In this
case, the central inversion is part of the rotation group, while the
specular reflection stays an extra operation. It turns out that the
symmetry we are really concerned with here (and which enters the CPT
theorem, with a corresponding definition of C) is the latter.

A somewhat surprising fact is that, in 4+1 dimensions, the simple
Lagrangian:
\begin{equation}\label{lagr5D}
    \mathcal{L}= i\bar{\psi}D \!\!\!\!/\psi -M \, \bar{\psi} \psi
\end{equation}
where $D^a= d/dx^a + i A^a$  is already P-violating. This is
unfamiliar, and will need a word of explanation, which will prove
the key to the CP breaking mechanism.

If we go back to 3+1 dimensions, we note that the two expressions
\begin{eqnarray}
  \bar{\psi}\psi&= &\bar{\psi}_L \psi_R + \bar{\psi}_R \psi_L  \\
  \bar{\psi} i \gamma_5 \psi&= &i (\bar{\psi}_L \psi_R - \bar{\psi}_R \psi_L)
\end{eqnarray}
are related by a chiral transformation, or more simply by the
allowed change
\begin{equation}\label{chiral}
   \psi_R \rightarrow i \psi_R
\end{equation}
There is no fundamental difference here between scalar and
pseudoscalr couplings, and the simultaneous presence of both
"scalar" and "pseudoscalar" independent terms is requested to have
parity violation in (3+1) dim.

This is not the case in 4+1 dimensions, since we can no longer
consider independently $\psi_L$, and $\psi_R$, but need to deal with
the full 4-component spinor grouping both. Seen otherwise, the
presence of $\gamma_4= i \gamma_5$ in the kinetic term effectively
forbids the transformation \eqref{chiral}.

\subsubsection{First example} Returning to \eqref{lagr5D}, we note
that afer singling out one extra dimension (typically, $x_4$), we
will generate terms of the type:
\begin{equation}\label{naive}
    \bar{\psi}(M+i \gamma_5 X_4) \psi
\end{equation}
where $X_4$ may originate either from the derivative or the $A_4$
term.

Such terms lead directly to a complex mass, even if the original
scalar couplings (M) were all real.

Now, of course two objections are in order. The first deals with the
observability of this complex coupling, namely would it imply CP
violation in 3+1 dimensions? We know the answer to this question is
negative, as a chiral transformation can be used to rotate the phase
away. In presence of other interactions, like QCD, this would
however contribute to the $\theta$ term. We will return to this in
the next section, as the obvious answer is to enlarge the gauge
group we consider (what we have here corresponds only to QED!). The
second objection deals with the gauge invariance of the procedure.
If indeed $X_4$ is related to the vacuum expectation value of a
field, like $X_4 = <A_4>$, the statement is not gauge invariant.

The answer to the second objection is of course straightforward, and
one has to introduce either the line integral (for unbounded space)
or the loop integral (for compact space):
\begin{equation}\label{line}
    X_4 = \int dy A_4 (y)
\end{equation}
where $y$ is the extra dimension. In the case of a closed loop, this
is just a Wilson loop. It can be thought of (if we imbed the loop in
an unphysical plane) as the flux of $A$ across the (unphysical)
surface spanned by the loop.

Alternatively, if we are dealing with a segment or orbifold
structure, $A_4$ can still be gauged away, but at the cost of
introducing non-(anti-)periodical boundary conditions induced by:
\begin{equation}\label{nonperiodical}
   \psi'(y)= e^{-i \int_0^y dy A_4(y) } \psi(y)
\end{equation}

The use of such a line integral to introduce symmetry breaking has
been investigated extensively by Hosotani \cite{Hosotani:1983xw},
\cite{Hosotani:1983vn}in the framework of dynamical symmetry
breaking.

\subsection{A working example}
We now turn to the simplest possible case of a working CP violation
model along the lines sketched above. For this purpose, we use a
fermion doublet, interacting with an $SU(2)$ gauge group according
to :
\begin{equation}\label{lagrSU2}
    \bar{\Psi}i (\partial^A -i W^A_a \tau^a) \gamma_A \Psi + M \bar{\Psi}\Psi
\end{equation}
and assume the Hosotani loop (seen here as a kind of boundary
condition):
\begin{equation}\label{loop}
    \langle W_4 \rangle =\int dy \; W_4(y) =\left(\begin{array}{cc} w&  \\  &-w \end{array}\right)
\end{equation}

This generates a mass matrix $$\mathcal{M}= \left(\begin{array}{cc}
M+iw\gamma_5&  \\  &M-iw\gamma_5\end{array}\right).$$ The phases can
be rotated away by a gauge transformation, but this leads to complex
coefficients at the charged $W$ vertices, leading to an effective
"$W_3$-dipole moment" through the graph shown in figure 2.
\begin{figure}
   \begin{picture}(200,110)(0,0)
      \ArrowLine(35,60)(65,60)
      \Text(50,80)[]{$e^{2i\alpha}$}
      \Vertex(65,60){2}
      \Line(65,60)(155,60)
      \Text(65,45)[]{$L$}
      \Vertex(155,60){2}
      \Text(155,45)[]{$R$}
      \Text(135,60)[]{$\times$}
        \Text(135,50)[]{$m$}
      \ArrowLine(155,60)(185,60)
      \PhotonArc(110,60)(45,0,180) {-4}{8}
        \ArrowArcn(110,60)(35,120,60)
      \Text(110,110)[b]{$W_{\pm}$}
      \Vertex(110,60){2}
      \Photon(110,0)(110,60){-4}{5}
      \Text(130,15)[l]{$W_3$}
    \end{picture}
        \caption{example of induced edm}
    \label{fig:edm}
\end{figure}
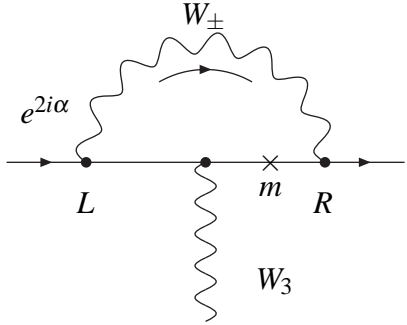

\subsection{Towards realistic models - chiral examples}
The above models have shown how to reach effective CP violation, but
are still a far cry from being realistic. For one thing, we have not
discussed how to obtain chiral fermions: indeed spinors in 4+1
dimensions contain both chiralities when naively reduced to 3+1
dimensions. The obvious trick is to use a topological defect (like a
domain wall, for instance). In this case, the index theorem shows
that for each 4+1-d fermion, we get only one localized chiral mode
in 3+1; the chirality (L or R) depending on the sign of the coupling
between the scalar representing the domain wall (the presence of the
latter is ensured by suitable boundary conditions). By choosing the
scalar responsible for the topological singularity to be part of a
larger gauge group, we can in fact obtain both chiralities from a
single representation: imagine for instance an octet of such
background scalars under SU(3) breaking with vev in the direction
$\lambda_8$, and coupled to a triplet of fermions.

The details of such mechanism become quite intricate, and can be
found in \cite{Cosme:2002zv} where we propose semi-realistic models,
endowed with non-trivial mass matrices and chiral fermions.

A complete picture finds its natural expression in the context of
grand unified theories. In this case, we have shown that the first
usable unification group is $SO(11)$ (rather than the usual $SU(5)$
or $SO(10)$). The source of this extension is precisely in the need
to localize both left- and right-handed fermions.

\section{Conclusion}

While I did not want to go into the technical details, I hope this
presentation has shown interesting and novel ways to address
fundamental questions like the family replication and the origin of
CP violation in the context of extra dimensions.


\begin{theacknowledgments}
   I wish to thank very warmly my colleagues Serguey Troitsky, Maxim
   Libanov and Emin Nugaev for our ongoing collaboration on the
   6-dimensional mode, as well as Nicolas Cosme and Laura
   Lopez-Honorez for the study of CP violation in extra dimensions.
   I thank of course the organisers of the CICHEP meeting in Cairo
   for a challenging conference, and acknowledge support from IISN
   and the belgian science policy (IAP V/27).
\end{theacknowledgments}

\bibliographystyle{aipproc}   

\end{document}